\documentclass{article}
\usepackage{spconf,amsmath,graphicx}
\usepackage{subcaption}
\usepackage{booktabs}
\usepackage{url}
\usepackage{hyperref}
\usepackage[utf8]{inputenc}
\title{Radiomic Feature Stability Analysis based on Probabilistic Segmentations}
\name{Christoph Haarburger\,$^{1}$, Justus Schock\,$^{2, 1}$, Daniel Truhn\,$^{3,1}$, Philippe Weitz\,$^{4}$,}
\secondlinename{Gustav Mueller-Franzes\,$^{1}$, Leon Weninger\,$^{1}$, Dorit Merhof\,$^{1}$}
\address{$^{1}$Institute of Imaging and Computer Vision, RWTH Aachen University, Germany \\
$^{2}$Department of Diagnostic and Interventional Radiology, University Hospital Düsseldorf, Germany \\
$^{3}$Department of Diagnostic and Interventional Radiology, University Hospital Aachen, Germany\\
$^{4}$Department of Medical Epidemiology and Biostatistics, Karolinska Institutet, Stockholm, Sweden}

\begin{document}
%
\maketitle
\begin{abstract}
Identifying image features that are robust with respect to segmentation variability and domain shift is a tough challenge in radiomics.
So far, this problem has mainly been tackled in test-retest analyses.
In this work we analyze radiomics feature stability based on probabilistic segmentations.
Based on a public lung cancer dataset, we generate an arbitrary number of plausible segmentations using a Probabilistic U-Net.
From these segmentations, we extract a high number of plausible feature vectors for each lung tumor and analyze feature variance with respect to the segmentations.
Our results suggest that there are groups of radiomic features that are more (e.g. statistics features) and less (e.g. gray-level size zone matrix features) robust against segmentation variability.
Finally, we demonstrate that segmentation variance impacts the performance of a prognostic lung cancer survival model and propose a new and potentially more robust radiomics feature selection workflow.

\end{abstract}
\section{Introduction}
\label{sec:intro}

Radiomics has been increasingly applied to radio- and oncological image data~\cite{aerts_2014,kickingereder_2016,rosa_2019}.
One of the main problems with the use of radiomics lies in the curse of dimensionality:
Most studies are conducted on several hundred images, while thousands of image features are extracted.
Several problems arise from this setup:
Firstly, by extracting such high numbers of features, despite the use of feature selection algorithms, multicollinearity and overfitting may lead to limited reproducibility of radiomic signatures~\cite{welch_2018,berenguer_2018}.
While image feature definitions are increasingly standardized~\cite{zwanenburg_2016,griethuysen_2017}, other problems lie in differences in image reconstruction~\cite{meyer_2019a} and in how the segmentations based on which the features are extracted, are generated.
In contrast to many claims~\cite{gillies_2016, yip_2016}, most current radiomics pipelines are not truly quantitative because segmentations are performed by humans and are thus subject to inter- and intra-rater variability~\cite{joskowicz_2019}.
We hypothesize that this limits the validity and reproducibility of radiomic signatures, even when they are based on expert segmentations.


Zwanenburg et al.~\cite{zwanenburg_2019} have already assessed robustness of features with respect to image perturbations such as translation, rotation and noise addition.
Aerts et al.~\cite{aerts_2014} and Peerlings et al.~\cite{peerlings_2019} investigated feature stability in test-retest studies.
In \cite{kalpathy-cramer_2016}, robustness of radiomics features with respect to domain shift was assessed across several institutions.
Owens et al.~\cite{owens_2018} evaluated uncertainty of radiomics features for manual versus semi-automatic segmentations.
In a recent review article~\cite{traverso_2018}, Traverso et al.~concluded that there is currently no consensus regarding the question which features are optimal in terms of reproducibility.
In order to quantify the degree to which radiomics features depend on the particular segmentation, we propose to perform a \textit{probabilistic} automated segmentation that generates a \textit{set} of plausible segmentations rather than one or few manual segmentations by experts.
Feature vectors are then extracted for each segmentation indidually, in order to assess robustness of individual features with respect to segmentation variability.

The set of plausible segmentations is generated by an extension of the U-Net~\cite{ronneberger_2015}, the Probabilistic U-Net(PU-Net)~\cite{kohl_2018}, which combines the U-Net with a conditional variational autoencoder (CVAE).
With this architecture, plausible segmentations can be sampled from the latent space of the CVAE.

Based on these, we extract a high number of radiomics features and assess feature variance with respect to a set of plausible segmentation masks.
We identify groups of features that are invariant with respect to the particular segmentation and others that depend on it more heavily.
Furthermore, we show that segmentation variance influences the performance of a prognostic survival model on a public lung cancer dataset.

\section{Material and Methods}
\subsection{Image Data}
\begin{figure*}[h]
  \centering
	\begin{subfigure}[t]{1in}
		\centering
		\includegraphics[width=1.1in]{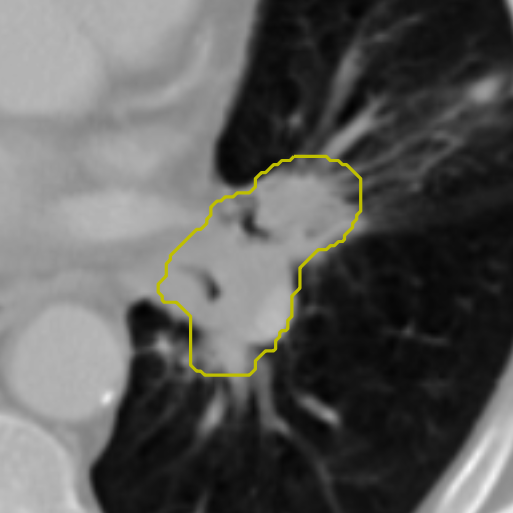}
		\caption{Ground Truth}\label{fig:1a}
	\end{subfigure}
	\quad
	\begin{subfigure}[t]{1in}
		\centering
		\includegraphics[width=1.1in]{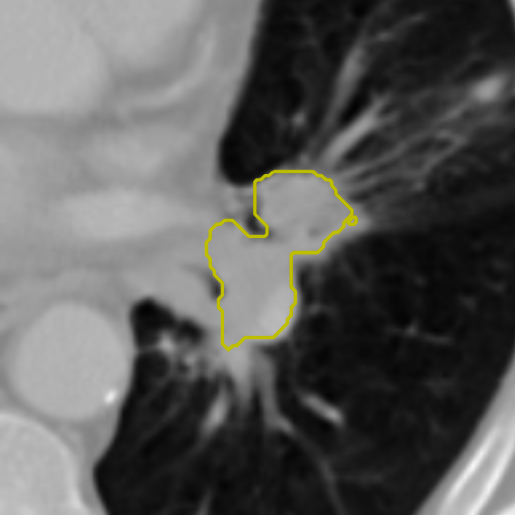}
		\caption{Segmentation \(D = 0.71\)}\label{fig:1b}
  \end{subfigure}
  \quad
  \begin{subfigure}[t]{1in}
		\centering
		\includegraphics[width=1.1in]{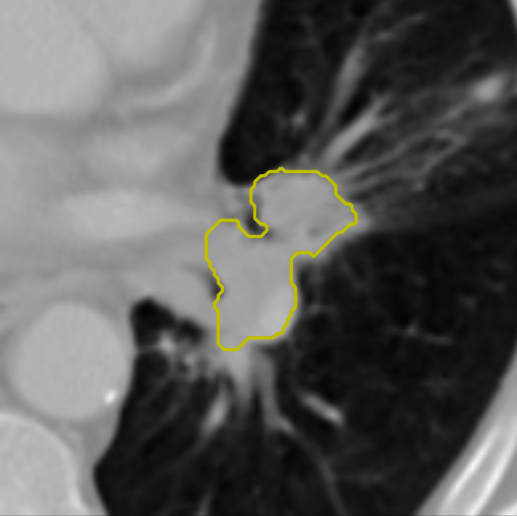}
		\caption{Segmentation \(D = 0.73\)}\label{fig:1c}
  \end{subfigure}
  \quad
  \begin{subfigure}[t]{1in}
		\centering
		\includegraphics[width=1.1in]{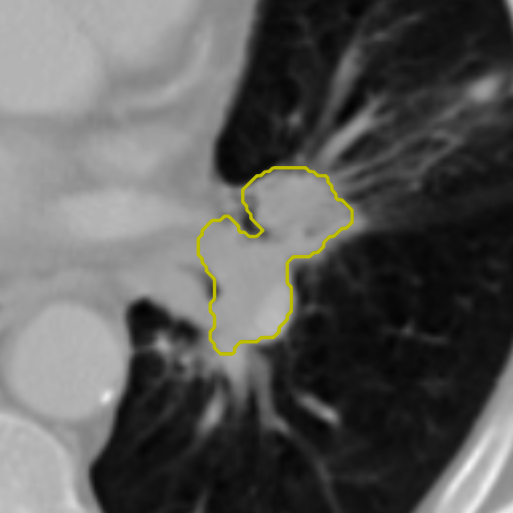}
		\caption{Segmentation \(D = 0.76\)}\label{fig:1d}
  \end{subfigure}
  \quad
  \begin{subfigure}[t]{1in}
		\centering
		\includegraphics[width=1.1in]{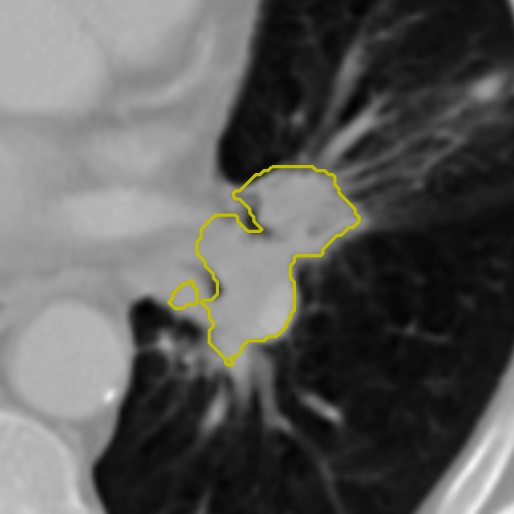}
		\caption{Segmentation \(D = 0.81\)}\label{fig:1e}
  \end{subfigure}
  \quad
  \begin{subfigure}[t]{1in}
		\centering
		\includegraphics[width=1.1in]{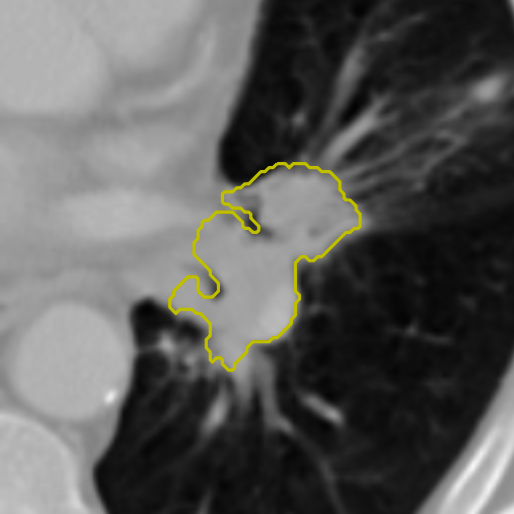}
		\caption{Segmentation \(D = 0.85\)}\label{fig:1f}
	\end{subfigure}
	\caption{Ground truth (a) and examples for corresponding probabilistic segmentations (b -- f). The dice scores \(D\) refer to the ground truth.}\label{fig:1}
\label{fig:lung1_example}
\end{figure*}

All evaluations are performed based on two publicly available lung cancer datasets:
Specifically, the \textit{LIDC-IDRI} dataset~\cite{armato_2011,armato_2015} is used to train the PU-Net, whereas feature stability is assessed on the \textit{Maastro Lung1} dataset~\cite{aerts_2014,aerts_2015}, which are both publicly available at \textit{The Cancer Imaging Archive (TCIA)}~\cite{clark_2013a}.
Expert segmentations for all 422 cases of the \textit{Lung1} dataset are also publicly available\footnote{\url{https://xnat.bmia.nl}}.
An example image with a corresponding expert segmentation is depicted in Fig.~\ref{fig:1a}.
Given the CT scans, expert annotations and right-censored survival data, the \textit{Maastro Lung1} dataset has been used for radiomics-based lung cancer \textit{survival analysis}~\cite{aerts_2014,welch_2018}.

\subsection{Segmentation using Probabilistic U-Net}
The PU-Net~\cite{kohl_2018} is an extension of the popular U-Net architecture~\cite{ronneberger_2015} that models the distribution of plausible segmentations using a CVAE.
By sampling from a low-dimensional latent space vector, plausible segmentation hypotheses can be generated arbitrarily which can be interpreted as an equivalent of asking a human experts to perform a manual segmentations.
We trained the PU-Net as implemented in~\footnote{\url{https://github.com/stefanknegt/Probabilistic-Unet-Pytorch}} using a latent vector with \(N=6\) as suggested in the original PU-Net publication~\cite{kohl_2018}.
The network operated on 2D axial slices.

After training the PU-Net on the \textit{LIDC-IDRI} dataset, 1000 plausible 2D segmentation masks were generated for every axial slice from the \textit{Maastro Lung1} dataset.
Unfortunately, for 73 cases, the probabilistic segmentation failed on several slices producing no segmentation at all. 
By visual inspection we were not able to identify any pattern signifying the failure of generating a segmentation.
These cases were excluded from this study.

\subsection{Feature Extraction}
\label{sec:featexts}
Using the PU-Net, we generated up to 1000 plausible segmentations for each slice.
This initial set of segmentations was then reduced to a set of \textit{unique} segmentations resulting in dozens to few hundred segmentations.
For this reduced set, we calculated the Dice score \(D\) between each probabilistic segmentation and the ground truth and sampled 25 segmentations with a uniform distribution with respect to the Dice scores related to the ground truth.
This step is needed to ensure that the segmentations actually differ because even in the reduced set of segmentations generated by the PU-Net, many are almost identical.
The particular choice of 25 segmentations was chosen based on visual inspection considering segmentation diversity and was not optimized further.

Next, for each nodule, 25 feature vectors, based on the 25 segmentations, were extracted using the PyRadiomics framework~\cite{griethuysen_2017} (Version 2.2.0).
Specifically, we extracted:
\begin{itemize}
  \itemsep-0.3em
  \item 18 statistics features
  \item 15 shape features
  \item 22 gray level co-occurence matrix (GLCM) features
  \item 16 gray level size zone matrix (GLSZM) features
  \item 16 gray level run length matrix (GLRLM) features
  \item 5 neighborhood difference gray tone matrix (NDGTM) features
\end{itemize}
For feature extraction, all images were resampled to a voxel spacing of 1\(\times\)1\(\times\)1\,\(\mathrm{mm^3}\) and a bin width of 25 was used for grey value binning.
Moreover, we extracted not only the image features based on original CT images but also on wavelet-transformed using the standard wavelets transforms implemented in PyRadiomics.

\section{Results}
For 73 out of the 422 patients, the PU-Net did not produce segmentation hypotheses as explained in Section~\ref{sec:featexts} but produced empty masks, which lead to an exclusion of these cases from our study.

\begin{figure}[]
    \centering
    \includegraphics[width=0.5\textwidth]{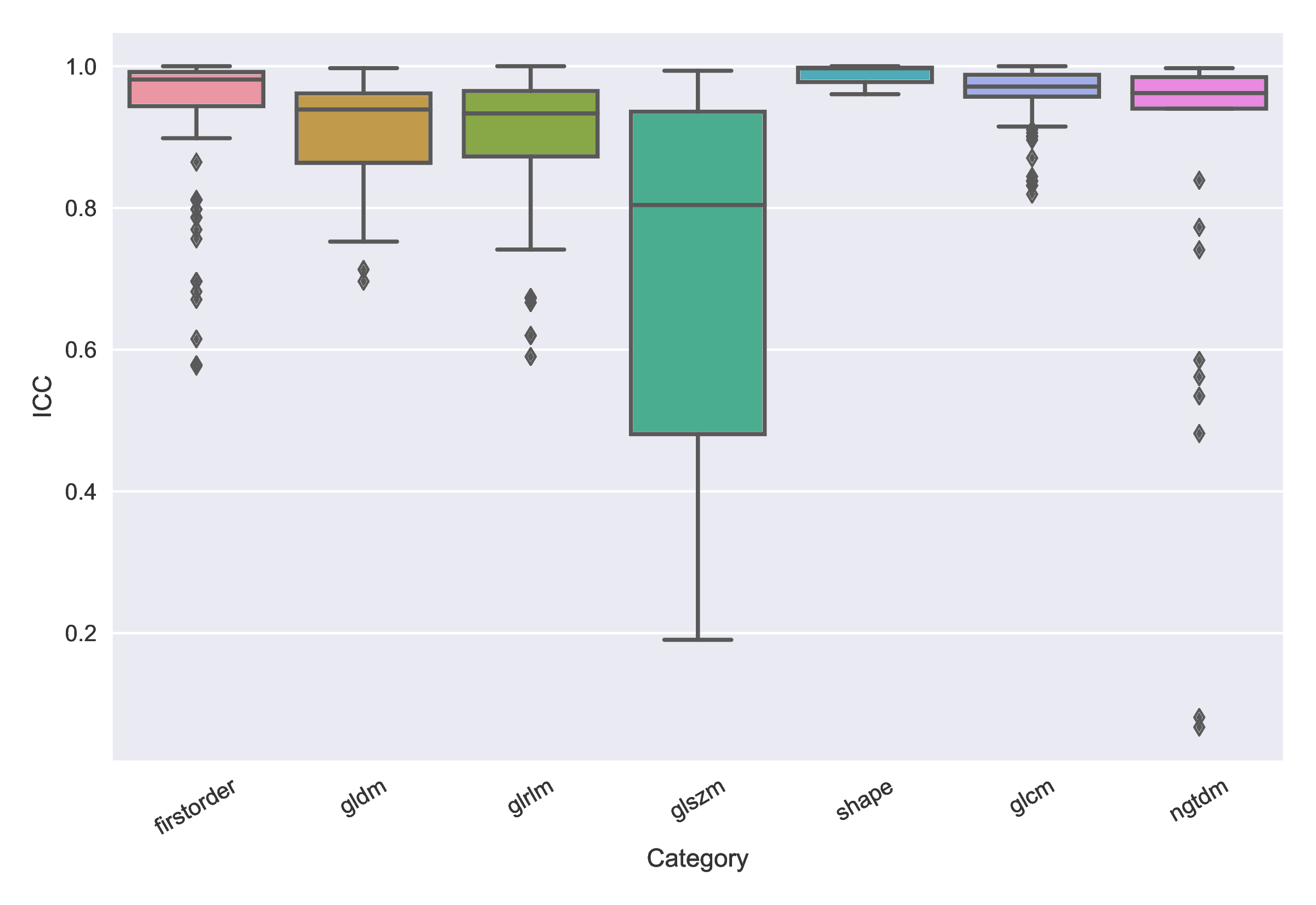}
    \caption{ICCs for all features clustered by feature category.}
    \label{fig:icc_category}
  \end{figure}

\begin{figure}[]
    \centering
    \includegraphics[width=0.5\textwidth]{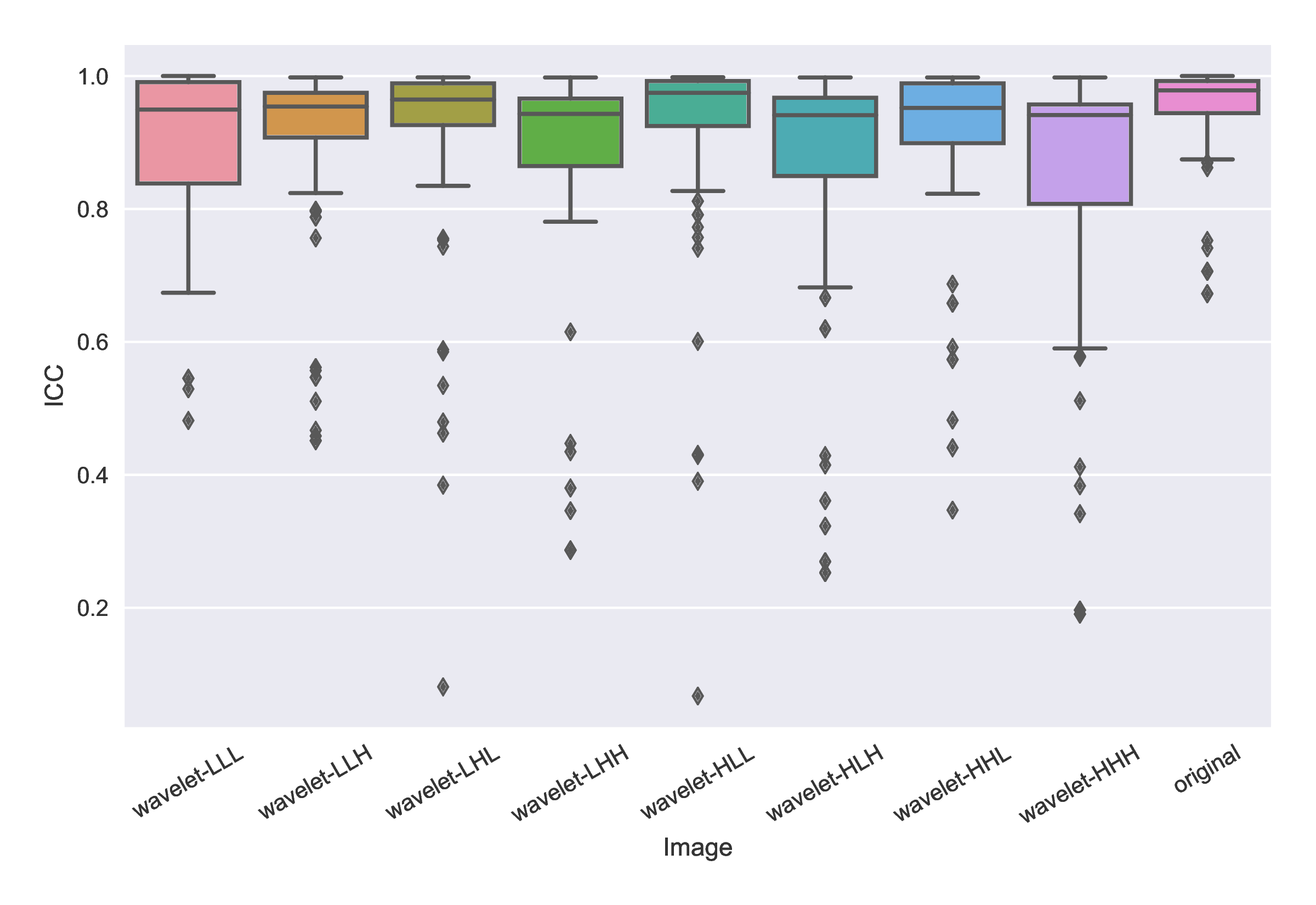}
    \caption{ICCs for all features clustered by image transform.}
    \label{fig:icc_image}
  \end{figure}

Feature stability was assessed by calculating the intraclass correlation coefficient (ICC) for each feature across all 25 segmentations.
ICCs grouped by feature categories are provided in Fig.~\ref{fig:icc_category}.
In Fig~\ref{fig:icc_image}, we provide the same ICCs clustered by image transform (no transform, wavelet transforms).

In Fig.~\ref{fig:icc_category} we can see that most statistics, shape, GLCM and NGTDM features have a very high ICC \(>\) 0.9.
Many GLRLM and especially GLSZM features have a considerably lower ICCs that can be as low as 0.2.
If the features are grouped by image transform as in Fig.~\ref{fig:icc_category} it becomes obvious that wavelet features are generally subject to higher segmentation dependence than features extracted from raw CT images.
Overall, 28.7\% of all features had an ICC \(<\) 0.9.

\begin{figure}[]
  \centering
  \includegraphics[width=0.5\textwidth]{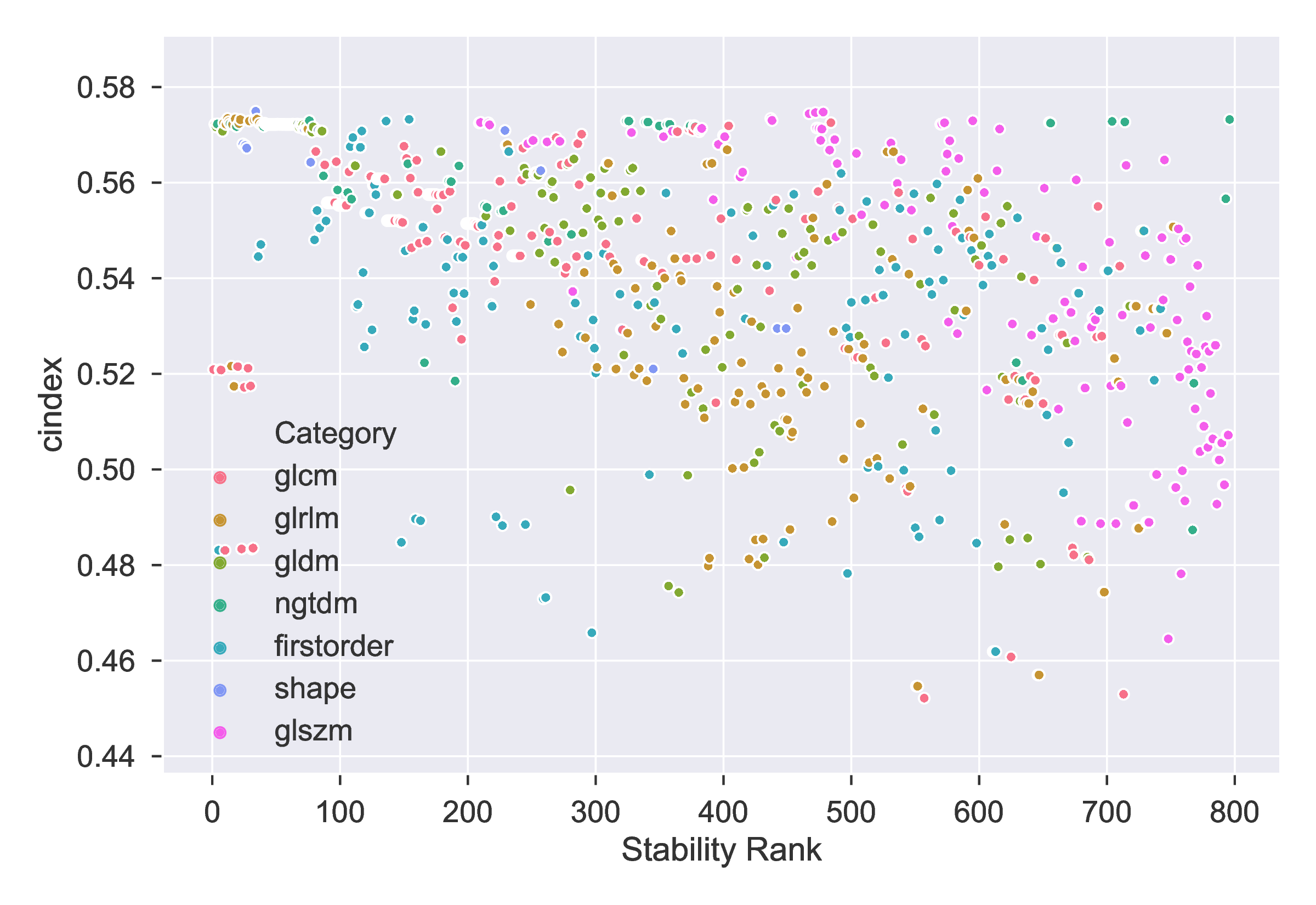}
  \caption{Stability rank (x-axis) and corresponding concordance index (cindex) for all features.}
  \label{fig:icc_stabilityrank}
\end{figure}

\begin{figure}[]
  \centering
  \includegraphics[width=0.5\textwidth]{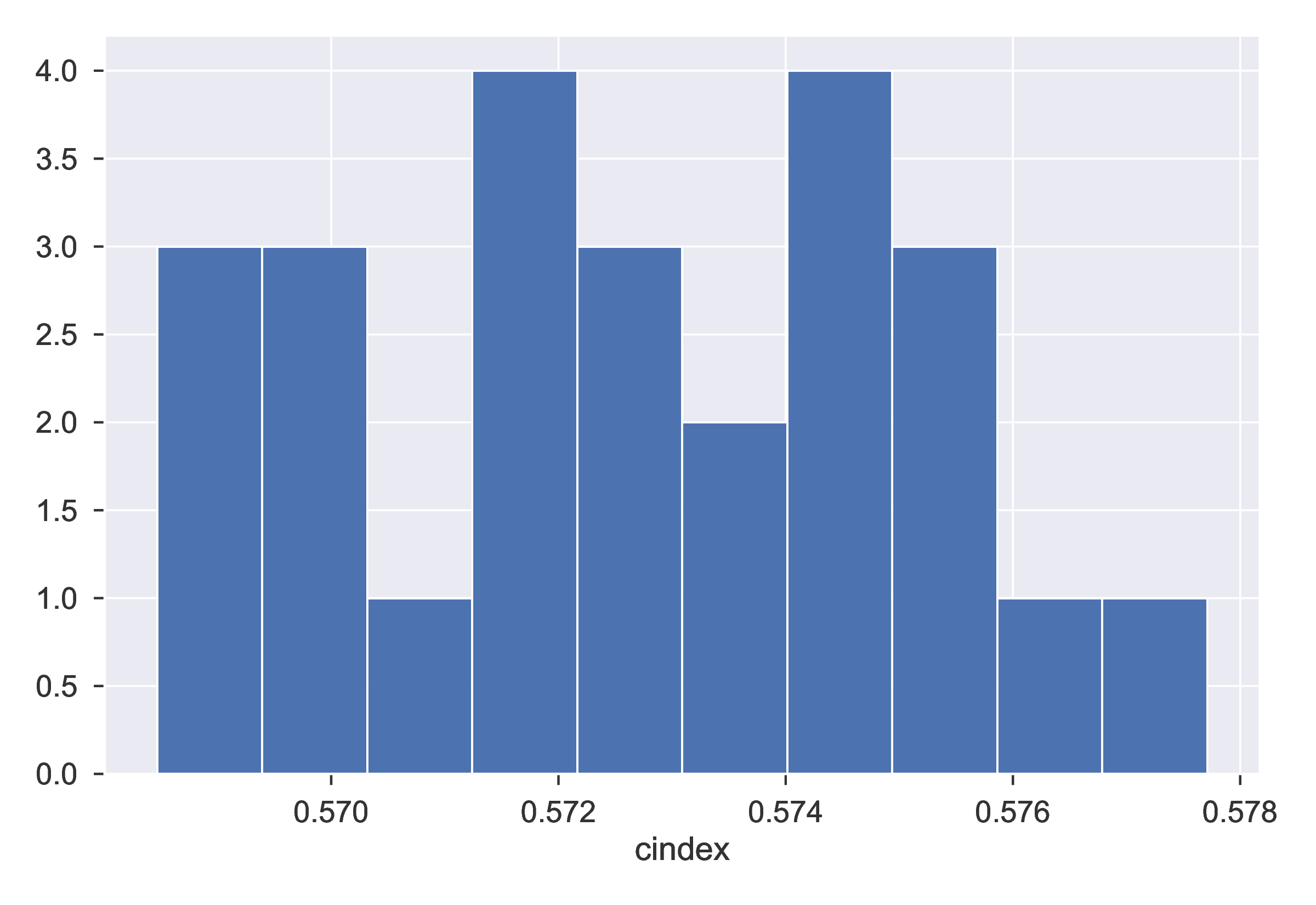}
  \caption{Histogram of cindices using a Cox model based on the Aerts signature~\cite{aerts_2014} over 25 probabilistic segmentations.}
  \label{fig:cindex_hist}
\end{figure}

In order to demonstrate the relation between the prognostic value and the stability a feature, in Fig~\ref{fig:icc_stabilityrank}, each for each feature the univariate concordance index (cindex) is plotted against the stability rank.
Cindex is a widely-used performance measure in survival analysis and defined as
\begin{equation}
\label{eqn:c_index}
\text{cindex}= \frac{\# \; \text{concordant pairs}}{\# \; \text{possible pairs}} \in [0,1].
\end{equation}
Stability rank is defined as a descending ranking of all features based on ICC such that the feature with the highest ICC has a stability rank of 1.
The last experiment is based on the radiomic signature by Aerts et al. based on a Cox model and the four features identified in~\cite{aerts_2014}.
We extracted radiomic signatures based on this model for all 25 segmentations and calculated the corresponding cindices, which are depicted in Fig.~\ref{fig:cindex_hist} as a histogram.
Cindices vary between 0.569 and 0.577.
In comparision, using the expert segmentation, a cindex of 0.574 is achieved.

\section{Discussion}

We have trained a probabilistic segmentation algorithm that provides segmentation hypotheses that can be used for extracting radiomics features.
Based on the feature vectors, we have analyzed feature stability with respect to varying segmentation masks.
The results show that there are groups of radiomics features that are subject to higher and lower variance across segmentations, respectively.
This is in line with other works in which stability of feature vectors originating from multiple manual segmentations by experts were evaluated~\cite{aerts_2014,peerlings_2019}.
Moreover, we were able to show in Fig~\ref{fig:icc_stabilityrank} that variance in segmentations carries over to a prognostic model.
However, the variance with respect to the cindex across segmentations is relatively small, indicating that the signature by Aerts et al.~is relatively robust against segmentation variance.

There is currently no consensus on which ICC cutoff should be used to exclude features from further analyses~\cite{traverso_2018}, however, assuming a cutoff of 0.9, about one third of all features in our analysis could be discarded.
Thus, in a standard pipeline for radiomic signature development, the curse of dimensionality and multicollinearity could be considerably alleviated, which may lead to radiomic signatures that are more robust and reproducible.
Moreover, feature scores could be averaged over segmentation mask hypotheses to further improve robustness.

Based on these findings, rather than ``just'' extracting radiomics features from a single expert segmentation, we envision that a future radiomic signature development pipeline could be composed of the following steps:
\begin{enumerate}
  \itemsep0em
  \item Train a probabilistic segmentation algorithm based on the expert segmentation
  \item Generate \(N\) plausible segmentations for each case
  \item For each feature, calculate ICC with respect to the \(N\) segmentations
  \item Discard features that are subject to low ICC
  \item For the remaining features, average feature scores over \(N\) probabilistic segmentations
  \item Run ``standard'' radiomics pipeline (feature selection, model fit, etc.)
\end{enumerate}

Our study has several limitations:
First, the dataset used for feature analysis originates from a single scanner, using a single reconstruction algorithm.
Thus, variability arising from protocol and device difference as investigated in~\cite{berenguer_2018} is not considered here.
Furthermore, the publicly available expert segmentations are prone to errors which might carry over to the generated probabilistic segmentations.
Accordingly, the PU-Net segmentations have several shortcomings:
First, segmentation is only performed on axial slices in 2D.
Moreover, training was partly unstable, producing many similar segmentation hypotheses that had to be filtered out.
Finally, for 73 cases, the automatic segmentation using the PU-Net failed completely producing empty masks.
These cases had to be excluded from our study and limit the comparability to other works on the same dataset.
Finally, our findings are only based on a single dataset from a single modality.
In future work, it is desirable to conduct the same experiments on other cancer types and imaging devices to assess feature robustness in a more general sense.

\section{Conclusion}
Using a set of plausible segmentation hypotheses generated by a PU-Net segmentation algorithm, we analyzed variance of radiomic features with respect varying segmentations, showing that there are groups of image features that are subject to different degrees of robustness.
Furthermore, we showed that segmentation variance carries impacts a radiomics survival model on a public lung cancer dataset.

\section{Acknowledgements}
The authors wish to acknowledge financial support from Interreg V-A Euregio Meuse-Rhine (“Euradiomics”).

\bibliographystyle{IEEEbib}
\bibliography{literature}

\end{document}